\documentclass[11pt]{article}
\input{epsf}
\begin{document}
\title{\bf Black holes as mirrors: \\quantum information in random subsystems}
\author{Patrick Hayden\\
{\em \small School of Computer Science, McGill University, Montreal, Quebec, H3A 2A7, Canada}\\ \\
John Preskill\\
{\em \small Institute for Quantum Information, California Institute of Technology, Pasadena CA 91125, USA}}
\date{}
\maketitle

\begin{abstract}
We study information retrieval from evaporating black holes, assuming that the internal dynamics of a black hole is unitary and rapidly mixing, and assuming that the retriever has unlimited control over the emitted Hawking radiation. If the evaporation of the black hole has already proceeded past the ``half-way'' point, where half of the initial entropy has been radiated away, then additional quantum information deposited in the black hole is revealed in the Hawking radiation very rapidly. Information deposited prior to the half-way point remains concealed until the half-way point, and then emerges quickly. These conclusions hold because typical local quantum circuits are efficient encoders for quantum error-correcting codes that nearly achieve the capacity of the quantum erasure channel. Our estimate of a black hole's information retention time, based on speculative dynamical assumptions, is just barely compatible with the black hole complementarity hypothesis.

\end{abstract}

\section{Introduction}
Is the information consumed by a black hole destroyed and lost forever \cite{hawk2}, or might it be recovered from the Hawking radiation that is emitted as the black hole evaporates? Evidence from string theory suggests that the information, rather than being destroyed, can be encoded in the black hole's internal degrees of freedom and eventually transferred to the outgoing radiation \cite{strominger-vafa,ads-cft}. However, the issue remains controversial, and in any event the mechanism by which information escapes from a black hole remains elusive.

Quantum information theory addresses quantitative questions about the acquisition, transmission, and processing of information in quantum systems \cite{bennett-shor}. Though quantum information theory cannot by itself resolve the black hole information puzzle, it can provide intuition and tools that help to sharpen our understanding of the question. 

In this paper, we assume that black holes, like other thermal systems, process quantum information rather than destroy it, and we apply insights from quantum information theory to study the information content of the Hawking radiation. Our conclusion is that, under plausible dynamical assumptions, the black hole releases information remarkably quickly, much faster than might have been naively expected.

Our analysis has two main components. At first, we assume that a black hole thermalizes quantum information arbitrarily quickly, so that we may model the internal dynamics of a black hole by an instantaneous random unitary transformation.  Under this assumption, we show in Sec.~\ref{sec:quantum} that if a black hole's internal degrees of freedom are nearly maximally entangled with the previously emitted Hawking radiation (as would be expected for a black hole that has already radiated away more than half of its initial entropy), then $k$ qubits of quantum information dumped into the black hole will be revealed after just a few more than $k$ qubits are emitted in the Hawking radiation. This observation rests on known achievable rates for entanglement-assisted quantum communication through a quantum erasure channel \cite{thapliyal}. 

Then we reexamine the issue of a black hole's thermalization time, and we argue in Sec.~\ref{sec:thermalization} and Sec.~\ref{sec:efficiency} that a black hole's internal quantum state becomes thoroughly mixed in a (Schwarzschild) time of order $r_S\log(r_S/l_{P})$, where $r_S$ is the black hole's Schwarzschild radius and $l_{P}$ is the Planck length (and where the speed of light is $c=1$). This argument, based on speculative dynamical assumptions, relies on a recent construction of efficient quantum circuits that realize approximate unitary 2-designs \cite{cleve,dankert}. Combining with the preceding result, we infer that, for a black hole whose evaporation is past the half-way point, $k$ qubits absorbed by the black hole will be reemitted in Schwarzschild time $O(kr_S)$ or $O(r_S\log(r_S/l_P))$, whichever is larger.

If we accept that black holes evolve unitarily and encode quantum information in their Hawking radiation, then we are faced with the challenge of reconciling this phenomenon with the perspective of an infalling observer who tumbles through the event horizon. We do not attempt to resolve this mystery here. Rather, we focus on the behavior of the black hole from the perspective of observers who stay outside. To these observers, a black hole is a seething cauldron of microscopic degrees of freedom localized close to the horizon, about one qubit per Planck unit of area, undergoing local unitary dynamics with a characteristic time scale of order the Planck time \cite{complement,susskind-book}. We assume that the observers refrain from attempting to probe these microscopic degrees of freedom directly, which would be far too dangerous. Rather they are content to infer how the black hole processes information indirectly, by investigating the relationship between the infalling matter and the outgoing radiation.

We will, however, address in Sec.~\ref{sec:cloning} whether our claim that information escapes rapidly from black holes can be reconciled with the hypothesis of ``black hole complementarity,'' according to which no violations of the accepted principles of quantum physics can be detected by any observer, whether outside or inside the black hole. We conclude that rapid escape and black hole complementarity are compatible, but only just barely so.

\section{A classical randomizer}
\label{sec:classical}
Black holes may not destroy information, but surely they hide it pretty well. How well?

The black hole information puzzle really concerns the processing of {\em quantum} information, but let us to begin our discussion by considering the fate of classical information that enters a black hole. Suppose that Alice, a citizen of a highly advanced civilization in the distant future who has recorded her most private thoughts in a very confidential diary, has second thoughts and resolves to destroy her diary. How should she proceed? Bob, the top forensic scientist of Alice's era, has remarkable capabilities --- he can recover the contents of an erased hard disk, restore the shredded pages of a document, even reconstitute  burned pages from their ashes and smoke. Presumably, Alice's safest option is to toss her diary into a nearby large black hole. Eventually, the black hole will evaporate completely, encoding Alice's diary in the outgoing Hawking radiation where it might be decrypted by Bob. But evaporation of a large black hole is an extremely slow process --- Alice's secrets will be secure not for all eternity, but at least for many generations to come.

Or will they? Since we are for now discussing only classical information, let us adopt a highly unrealistic classical model of a black hole. (It will be instructive to contrast this classical model with a quantum model of a black hole that we will discuss in Sec.~\ref{sec:quantum}.) In this classical model, Alice's diary is a bit string of length $k$ and the internal state of the large black hole is regarded as a bit string of length $n-k \gg k$.  We assume that Bob, who has been observing the black hole since its formation and has a thorough understanding of black hole dynamics, knows the black hole's internal state, but he does not (yet) know the content of Alice's diary.

Now Alice tosses in her diary; the black hole's bit string grows to length $n$, where Bob knows $n-k$ of the bits, and the black hole's internal dynamics processes this length-$n$ string. We model this dynamics as a permutation (known by Bob) of all of the $2^n$ strings of length $n$ ({\em not} a permutation of the $n$ bits). After the processing, the black hole releases the bits one at a time in the Hawking radiation, as Bob watches expectantly. How soon will Bob be able to read the diary? We claim that, for almost any permutation, Bob will only need to receive a few more than $k$ bits before he will be able to decipher the complete diary, with low probability of error. Alice's secrets are {\em not} protected for the full lifetime of the black hole; rather they are revealed to Bob almost as quickly as possible! 

The black hole dynamics is deterministic (one particular known permutation is applied to the $n$-bit string), but for analyzing the information content of the radiation it is helpful to adopt the information theorist's favorite trick --- to assume that the permutation has been chosen uniformly at random from among the $\left(2^n\right)!$ possible ones. After processing by the black hole, Alice's $k$-bit message has been transformed to one of $2^k$ possible $n$-bit strings, and if Bob could read all $n$ bits he would know which of the $2^k$ strings he had and so decode Alice's message. But even if Bob has access to just a few more than the first $k$ bits of the string, he is likely to be able to rule out all messages except for the correct one, so that he can still decode successfully.

For the case of a random encoding, the probability of a decoding failure is easily estimated. If Bob reads the first $s$ bits of the string, what is the probability that these bits accidently match the first $s$ of an encoded message other than the correct one? For each message, the probability of an accidental match is $2^{-s}$, and since there are all together $2^k$ encoded messages, the probability $P_{\rm fail}$ that any of the wrong messages match the $s$ bits satisfies
\begin{equation}
\label{classical-out}
P_{\rm fail} \le 2^k 2^{-s}= 2^{-c}~,\quad {\rm where} \quad s=k+c~.
\end{equation}
Therefore, if Bob wants the probability of failure to be no larger than $2^{-c}$ for some constant $c$, he decodes after receiving $k+c$ bits of Hawking radiation. Here we speak of a {\em probability} of failure because we are averaging over all the possible encodings of $k$ bits in a block of $n$ bits. Our conclusion is that most encodings work. 

In information theory, the {\em capacity} $C$ of a noisy communication channel is the maximum achievable asymptotic rate at which coded information can be sent through the channel with a negligible probability of a decoding error by the receiver. What we have just described is related to two standard results in the theory of noisy classical channels \cite{cover}: (1) The (classical) {\em erasure channel} with erasure probability $p$ has capacity $C=1-p$, and (2) random encodings achieve the capacity.

In our setting, the black hole dynamics transforms Alice's $k$-bit message into one of the $2^k$ codewords of a random code with block length $n$. We say that the rate of the code is $R=k/n$, the number of message bits per bit in the code block. When $k+c$ of the bits in the block have been revealed to Bob, the remaining $n-k-c$ bits have in effect been erased as far as Bob is concerned. Yet, despite these erasures, Bob is able to decode Alice's message with good success probability. Letting $n$ get large with $R$ and $c$ fixed, we conclude that the message can be decoded even as the fraction of erased bits approaches $p=1-R$; that is, the rate $R=1-p$ is achievable in the limit of a large code block.

To Alice's dismay, we conclude that (in this model) a black hole is hardly black at all; it might more accurately be regarded as a kind of {\em information mirror}. Alice throws her diary into the black hole, and it bounces right back! Granted, it may be a strange sort of mirror, since if the Hawking radiation leaks out slowly and $k >> 1$, then Alice's message is obscured for a while; furthermore, Bob needs to use his knowledge of the black hole's initial state and its dynamics to decipher the ``reflection.'' But once a few more than $k$ bits have been reemitted by the black hole, the diary comes back into sharp focus, and Alice's secrets are no longer concealed from Bob. What is especially ironic about this scenario is that, by modeling the internal black hole dynamics as a random permutation, we hoped to maximize the black hole's power to hide the information it consumes. But at least as a matter of principle we have achieved the opposite of what we intended ---  the random permutation encodes the $k$ bits in a form that is optimally protected against the damaging effects of erasure! 

Let's point out some implicit assumptions underlying this model. We have assumed that the internal dynamics of the black hole is very fast --- the permutation is applied almost instantaneously after Alice's message is deposited, before any of the Hawking radiation leaks out. We will return to the issue of estimating the actual thermalization time scale in Sec.~\ref{sec:thermalization} and Sec.~\ref{sec:efficiency}. We have assumed that the permutation is ``typical.'' This assumption is nontrivial, because if the dynamics of the black hole is realized by a (reversible) classical computer performing local logic gates, then most permutations require very long computations, and the computations that can be performed {\em efficiently} may be far from typical. Similarly, we have not worried about the efficiency of Bob's decoding operation. We have imagined that Bob consults a huge codebook that lists the $2^k$ valid message strings, but if $k$ is large then this codebook would be of unmanageable size. We will discuss the efficiency of the recovery procedure further in Sec.~\ref{sec:efficiency}. Finally, the most glaring drawback of our model is that it is classical. Let us now turn to its quantum generalization.

\section{A quantum randomizer}
\label{sec:quantum}

Again, we imagine that Alice regrets recording some information and wants to destroy it, but this time the information is not a bit string --- rather it is {\em quantum information} stored in a $k$-qubit quantum memory. Normally, when we say that a quantum memory stores $k$ qubits, we mean that the stored quantum state lives in a Hilbert space of dimension $2^k$, but we also mean something more: that the Hilbert space has a physically natural decomposition as a tensor product of $k$ two-level systems. For example, we might envision the memory as a system of $k$ spin-$\frac{1}{2}$ particles. However, this tensor product decomposition will not be central to our discussion, so it will for the most part be adequate to regard Alice's message system $M$ as a Hilbert space of dimension $|M| =2^k$ without any special structure (and where $k$ need not be an integer).

Next, we need to reconsider some other features of the classical scenario. For example, what does it mean to say that Alice's quantum state can be recovered by Bob from the Hawking radiation? We don't necessarily mean that Bob can acquire a complete classical description of the state; that would be too much to ask. Rather we mean that Bob can do anything with the recovered state that he would have been able to do with the state of Alice's memory if he had been able to access it in the first place, before Alice tried to destroy it. 

It is useful to imagine that a third party (Charlie) holds a {\em reference system} $N$ with dimension $|N|=|M|$ that is maximally entangled with Alice's memory $M$. That is, the initial joint state of the memory and the reference system may be chosen to be the pure state
\begin{equation}
|\Phi\rangle^{MN}=\frac{1}{\sqrt{|M|}}\sum_{a=1}^{|M|} |a\rangle^M\otimes |a\rangle^N~;
\end{equation}
we say that the $N$ provides a {\em purification} of the state of $M$. If Charlie holds $N$, but Alice retains $M$, then the density operator for $N$ (upon tracing out $M$) is maximally mixed. Suppose that Alice tosses $M$ into the black hole. If sometime later Bob is able to extract from the Hawking radiation a subsystem of dimension $|M|$ that is maximally entangled with $N$, then we may say that Bob has recovered the quantum information that had been stored in Alice's quantum memory. This would imply in particular that if the initial state of $M$ had been a pure state $|\psi\rangle$ (not entangled with any reference system), then Bob would be able to recover $|\psi\rangle$ in his chosen subsystem \cite{barnum}.

In the classical scenario, we imagined that Bob knew the initial internal state of the black hole and the dynamical laws that govern its evolution. We would like to make parallel assumptions regarding the quantum model, but what should it mean to say that Bob ``knows'' a quantum state? When we say that Bob knows the bit string encoded in the classical black hole, we really mean that Bob holds a record that is perfectly correlated with the state of the black hole ({\em e.g.}, a perfect copy of the string). Similarly, in the case of the quantum black hole we may imagine that Bob holds a quantum memory that is perfectly correlated with the black hole's internal state, {\em i.e.}, maximally entangled with it. This is a strong assumption, but not a crazy one if we grant Bob complete control over the Hawking radiation. 

Indeed, consider how the entanglement of the black hole with the emitted radiation evolves as the black hole evaporates. We may divide the world into two subsystems --- the black hole's internal system $B$ and radiated system $E$. The relative size  of these subsystems varies with time; in particular, we may assume that, at any stage of the evaporation process, $\log |B|$ is the black hole's Bekenstein-Hawking entropy. Early on, soon after the black hole's formation, we have $|B|/|E|\gg 1$, and one can plausibly argue \cite{page-entropy,lubkin,lloyd} that $E$ is very nearly maximally entangled with a subsystem of $B$. However, as the evaporation proceeds,  $\log |B|$ eventually declines to half of its initial value, and soon after we have $|B|/|E| \ll 1$; then we may expect that $B$ is very nearly maximally entangled with a subsystem of $E$.  

Suppose that Alice, intent on destroying her $k$-qubit quantum memory, heads for the nearest large black hole and tosses her qubits in. In her haste, she imprudently fails to investigate the hole's history. But this particular black hole actually formed long ago, and Bob has been collecting its emitted Hawking radiation ever since. By now, the black hole's internal state is maximally entangled with a system Bob controls.

How soon will Bob be able to recover Alice's memory from the Hawking radiation? We assume that the internal dynamics of the black hole is a deterministic unitary transformation that thoroughly mixes the infalling information into the black hole's preexisting $(n-k)$-qubit state; then the black hole's qubits are released, one by one, in the Hawking radiation.  We claim that, for almost any unitary transformation,  Bob needs to wait for only a few more than $k$ qubits to be emitted. Much as in our classical discussion, the (maximally entangled) black hole is hardly black at all --- it is a quantum information mirror that returns to Bob the information Alice deposited almost as quickly as possible!

\begin{figure}
\begin{center}
\leavevmode
\epsfxsize=3.5in
\epsfbox{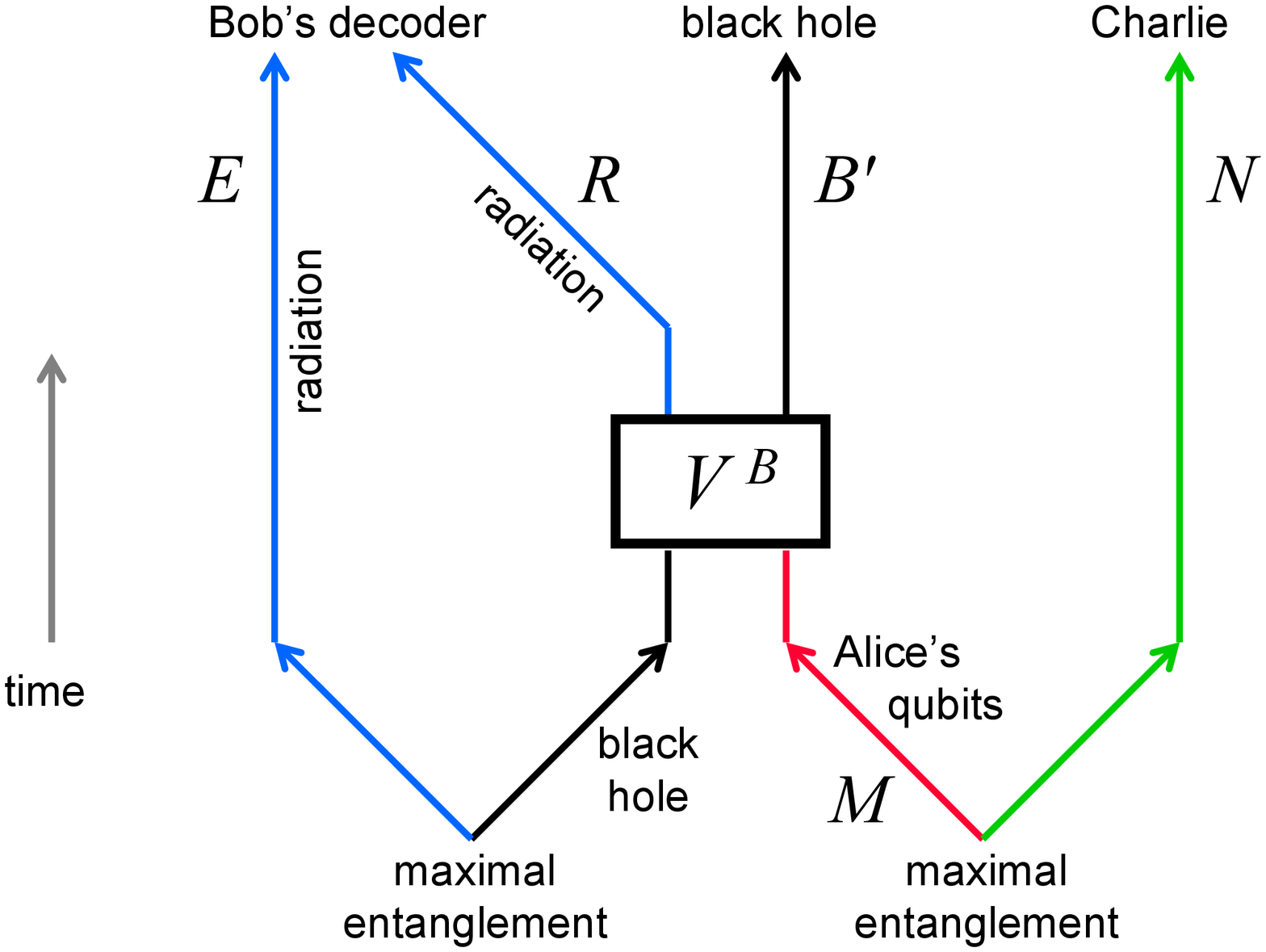}
\end{center}
\caption{Information retrieval from an evaporating black hole. The black hole, which has been evaporating for a long time, has become maximally entangled with the previously emitted radiation system $E$. At this stage, the black hole swallows Alice's quantum memory $M$, which is maximally entangled with Charlie's reference system $N$. The internal dynamics of the black hole applies a strongly mixing unitary transformation $V^B$, and then the additional radiation system $R$ is emitted, where the dimension of $R$ is somewhat larger than the dimension of $M$. Now a subsystem of $RE$, controlled by Bob, is nearly maximally entangled with $N$ --- the content of Alice's quantum memory has escaped from the black hole and is in Bob's possession.}
\label{fig:retrieval}
\end{figure}

Right after Alice tosses in her qubits, the $n$-qubit black hole system $B$ is maximally entangled with the system $NE$; here $B$ includes Alice's memory system $M$, which has now been absorbed by the black hole,  $E$ is the previously emitted Hawking radiation controlled by Bob, and $N$ is Charlie's reference system that had been entangled with $M$. As Bob watches attentively, the black hole continues to emit Hawking radiation until, after a while, $s$ additional qubits (the subsystem $R$ of $B$) have been emitted, with $n-s$ qubits (the subsystem $B'$) still retained by the black hole. We suppose for now that the emitted $s$-qubit subsystem $R$ of $B$ is chosen uniformly at random (we will revisit this assumption in Sec.~\ref{sec:efficiency}). That is, we imagine that  $B$ is divided into two parts, one with $s$ qubits and the other with $n-s$ qubits; then a unitary transformation $V^B$ chosen uniformly with respect to the Haar measure on $U(2^n)$ is applied to $B$. After that, the $s$-qubit system is identified as $R$.

As the Hawking radiation leaks out, the correlations between the evaporating black hole $B'$ and the reference system $N$ gradually weaken. Once $R$ is large enough, the surviving correlation of $N$ with $B'$ becomes negligible. At that point, since the overall state of $B'RNE$ is pure, the state of the reference system $N$ is very nearly purified by the radiation system $RE$ that Bob controls --- by this time, Alice's quantum information has fallen into Bob's hands. See Fig.~\ref{fig:retrieval}.

Let $\Psi^{BNE}$ denote the pure density operator of $BNE$,  and let $\rho^{BN}={\rm Tr}_E\Psi^{BNE}$ be the corresponding marginal density operator on $BN$. The marginal density operator on $NB'$ is
\begin{eqnarray}
\sigma^{NB'}(V^B)= {\rm Tr}_R \left[\rho^{NB}(V^B)\right]~, \quad {\rm where}\quad 
\rho^{NB}(V^B)=\left(I^N\otimes V^B\right)\rho^{NB}\left(I^N\otimes {V^B}^\dagger \right)~.
\end{eqnarray}
Using standard estimates \cite{abeyesinghe,yard}, we find that the $L^1$ distance of $\sigma^{NB'}$ from a product state, averaged over $V^B$ (and hence over the choice of the subsystem $R$), can be bounded as
\begin{eqnarray}
\int dV^B\left\| \sigma^{NB'}(V^B)-\sigma^{N}(V^B)\otimes \sigma^{B'}_{\rm max}\right\|_1^2 \le \frac{|NB|}{|R|^2}~{\rm Tr}\left[\left(\rho^{NB}\right)^2\right]~;
\end{eqnarray}
here $\sigma^{N}(V^B)={\rm Tr}_{B'}\left[\sigma^{NB'}(V^B)\right]$
is the marginal density operator on $N$, and $\sigma^{B'}_{\rm max}= I^{B'}/|B'|$ is the maximally mixed density operator on $B'$. The $L^1$ norm, defined by $\|A\|_1= {\rm Tr} \sqrt{A^\dagger A}$, is an appropriate measure because two states that are close in this norm cannot be well distinguished by any measurement \cite{fuchs}. 

In the case we are currently considering, in which $B$ is maximally entangled with $NE$, $\rho^{NB}$ is maximally mixed on a system of dimension $|B|/|N|$, and therefore 
\begin{equation}
{\rm Tr}\left[\left(\rho^{NB}\right)^2\right]~= |N|/|B|~;
\end{equation}
hence we find
\begin{eqnarray}
\label{quantum-out}
\int dV^B\left\| \sigma^{NB'}(V^B)-\sigma^{N}(V^B)\otimes \sigma^{B'}_{\rm max}\right\|_1^2 \le \frac{|N|^2}{|R|^2}= \frac{2^{2k}}{2^{2s}}= 2^{-2c}~,\quad {\rm where}\quad s=k+c~.
\end{eqnarray}
We see that, for a typical unitary transformation $V^B$ and for $c$ sufficiently large, the state of $NB'$ is nearly maximally mixed after $k+c$ qubits have escaped from the black hole. Alice's $k$ qubits have been ``forgotten'' by the black hole and have been acquired by Bob.  

Inconveniently, the information originally encoded in Alice's memory $M$ has become encoded in a subsystem $M'$ of $RE$ that is very diffusely distributed among the emitted radiation quanta. But in principle Bob could do a quantum computation that maps $M'$ to a compact system $\hat M$ localized in his laboratory. For any fixed value of the unitary transformation $V^B$, Bob's decoding map can be chosen so that, after decoding, the density operator $\rho^{\hat M N}$ of $\hat MN$ is close to the maximally entangled state $|\Phi\rangle^{\hat MN}$ (see, for example, \cite{yard}):
\begin{equation}
F(V^B) \equiv \langle \Phi|\rho^{\hat MN}|\Phi\rangle~\ge ~ 1-\left\| \sigma^{NB'}(V^B)-\sigma^{N}(V^B)\otimes \sigma^{B'}_{\rm max}\right\|_1.
\end{equation}
Eq.~(\ref{quantum-out}) implies that, after averaging over $V^B$, the fidelity $F(V^B)$ deviates from one by no more than $2^{-c}$; apart from a small error, Bob holds the purification of Charlie's reference system $N$. Thus eq.~(\ref{quantum-out}) can well be regarded as the quantum analog of the classical estimate eq.~(\ref{classical-out}); in both the classical and quantum models, the information that Alice deposits in the black hole escapes almost as soon as it possibly can. 

The conclusion that $\sigma^{NB'}$ becomes nearly maximally mixed after about $k$ randomly chosen qubits are discarded can be understood heuristically as follows: The initial density operator $\rho^{NB}$ is uniform on a space of dimension $|B|/|N|$, and so can be expressed as a uniform ensemble of $|B/|N|$ mutually orthogonal pure states.  If $|R| \ll |NB'|$, then a typical pure state on $NB$ is close to maximally mixed on $R$; therefore after tracing out $R$ each pure state in the ensemble representing $\rho^{NB}(V^B)$ yields a uniform density operator on a subspace with dimension $|R|$ of $NB'$. All together then, ignoring the overlap of these subspaces, the density operator $\sigma^{NB'}$ is nearly uniform on a space of dimension $|B|\cdot |R|/|N|$. This matches the dimension of $NB'$, $|N|\cdot |B'|=|N|\cdot |B|/|R|$, for $|R|=|N|$. We expect then, that for $|R|\approx |N|$, the density operator $\sigma^{NB'}$ is nearly uniform on $NB'$. 

As in the classical model, our conclusion can be usefully restated in terms of known results in the theory of noisy channels: for the {\em quantum erasure channel} with erasure probability $p$, the {\em entanglement-assisted quantum capacity} is $Q_E=1-p$ \cite{thapliyal}. In our setting, the $n-k-c$ qubits retained by the black hole are in effect erased as far as Bob is concerned, yet Bob is able to extract $k$ qubits of high fidelity quantum information. The quantum communication rate $R\approx k/n$ is achieved using a randomly chosen unitary encoder, and by exploiting a supply of pre-existing quantum entanglement that Bob shares with the black hole. 

Suppose on the other hand that Alice is more cautious, and deposits her sensitive quantum information in a relatively young black hole, such that $|E|/|B| \ll 1$. In that event, the previously emitted Hawking radiation $E$ will be nearly maximally entangled with a subsystem of $B$. The radiation will continue to be essentially featureless, revealing none of Alice's information, until $|B'|=|NRE|$. Soon after, the black hole will be nearly maximally entangled with its surroundings and eq.~(\ref{quantum-out}) will begin to apply; at that stage Alice's information suddenly spills out. 

Here too the conclusion is related to a well known property of the quantum erasure channel with erasure probability $p$: its quantum capacity (unassisted by entanglement) is $Q=0$ for $p\ge 1/2$ and $Q=1-2p$ for $p < 1/2$ \cite{quantum-erasure}. If the black hole initially holds $n$ qubits, then when $(n+k)/2$ qubits have emerged, $(n-k)/2$ qubits remain inside the black hole. At this point, Bob's system is $k$ qubits larger than the black hole, so that Bob has acquired $k$ qubits of quantum information.  Thus the communication rate is $R=k/n$ and the erasure probability is $p= 1/2 -k/2n=1/2 - R/2$.

Some years ago, Don Page observed that the information swallowed by a black hole remains hidden until half of the black hole's qubits have been radiated away, and then emerges at a constant rate as the evaporation proceeds \cite{page-entropy}. Our conclusion is similar, but our analysis goes further. Page considered the time dependence of the quantum entanglement of the black hole with its surroundings, which starts out small, grows until half of the entropy has been radiated away, and then declines. We have focused instead on when a fixed amount of quantum information of particular interest can be recovered.

Our simple model of quantum black holes leads us to two main conclusions, under the assumption that Bob has unlimited control over the Hawking radiation. If Alice dumps $k$ qubits into the black hole after half of the black hole's initial entropy has been radiated away, the information bounces right back --- Bob recovers Alice's quantum state after waiting for just a few more than $k$ qubits to be emitted. On the other hand, if Alice dumps $k$ qubits into the black hole before half of the black hole's initial entropy has been radiated away, then Bob needs to wait for a while. But once the black hole reaches the stage where half of its qubits have been released (and the black hole has become maximally entangled with its surroundings), Alice's information pops out almost immediately!

Both statements seem strange upon first hearing, but especially the latter one, because who is to say which $k$ qubits are ``Alice's''? In fact, no matter which $k$ qubits of quantum information swallowed by the black hole are of particular interest, these $k$ qubits are revealed almost right away when the half-way point of evaporation is finally reached. There is nothing special about the subsystem $M$ of $B$ that is maximally entangled with $N$ in Fig.~\ref{fig:retrieval}; for any other $k$-qubit subsystem the conclusion would have been the same --- that $N$ becomes very nearly maximally entangled with a $k$-qubit subsystem of $RE$. 

Therefore, when a black hole that initially held $n$ qubits has evaporated past the half-way point, so that $(n+k+c)/2$ qubits have been emitted, Bob gets to decide {\em which} $k$ qubits of quantum information he will retrieve from the Hawking radiation;  when he makes up his mind he performs the decoding operation on $RE$ that maps those $k$ qubits to the quantum memory in his laboratory. But the catch is that, although Bob can recover {\em almost any} $k$-qubit subsystem at this stage, he cannot recover more than $k$ qubits. 

Returning to the situation depicted in Fig.~\ref{fig:retrieval}, imagine two different $k$-qubit subsystems of the black hole, $M_1$ purified by reference system $N_1$ and $M_2$ purified by reference system $N_2$. When Bob decodes only $M_1$ his decoding map is permitted to act on $N_2$ (which in that case may be considered to be part of the radiation system $RE$), and when he decodes only $M_2$ his decoding map may act on $N_1$. But if he greedily wanted to decode both $M_1$ and $M_2$, then both $N_1$ and $N_2$ would be out of his reach; thus trying to decode $M_2$ in addition to $M_1$ only interferes with his ability to decode $M_1$. If Bob wants to retrieve more than $k$ qubits, he'll just have to wait for the black hole to emit more quanta.

We note that if the internal dynamics of the black hole is actually
time-reversal invariant, then it may be more appropriate to model the
dynamics by choosing the matrix $V^B$ from either the circular orthogonal
or the circular symplectic ensemble, with the choice depending on the
total spin of the black hole~\cite{mehta}.  Since time-reversal invariance
is not a fundamental symmetry in Nature, we think it is sensible to
consider $V^B$ chosen uniformly with respect to the Haar measure on
$U(2^n)$ as described above. We have in any case verified that our
conclusions would not be much affected were we to choose $V^B$ from the
circular orthogonal ensemble or circular symplectic ensemble instead \cite{anderson}.

\section{The thermalization time}
\label{sec:thermalization}
As for our classical model, there are some nontrivial assumptions underlying our model of a quantum black hole. In particular, when we conclude (in the case where the black hole is maximally entangled with previously emitted radiation) that Alice's information ``comes out fast,'' we are taking it for granted that the information deposited by Alice is rapidly thermalized by the black hole's internal dynamics. By rapidly, we mean on a time scale comparable to the time interval between the emission of successive radiation quanta. In terms of the ``Schwarzschild time'' measured by static observers who are far from the black hole, this interval is of order the Schwarzschild radius $r_S$ for a nonrotating uncharged black hole (in units where the speed of light is $c=1$).

There are several reasons why $r_S$ might seem at first to be a reasonable estimate of the black hole's thermalization time \cite{horowitz}. Classically, perturbed black holes have damped quasinormal ``ringing'' modes, where the damping time is comparable to $r_S$. Thus, if one kicks a classical black hole, $r_S$ is the time scale for the black hole to forget about the kick. Furthermore, under AdS/CFT duality, the dual of a large black hole is a strongly-coupled conformal field theory with the same temperature $T\approx 1/r_S$ (in units with $\hbar=c=1$). In the field theory the only relevant time scale governing the approach to thermal equilibrium is the time scale set by the temperature itself: $1/T\approx r_S$. Thus, if one kicks the strongly-coupled thermal bath, $r_S$ is the time scale for the bath to forget about the kick.

However, when one says that the strongly-coupled quantum field theory thermalizes in a time of order $1/T$, one ordinarily means that the system relaxes to a state that is {\em locally} indistinguishable from a thermal state on this time scale. Our assertion that information captured by a black hole is quickly reemitted in the Hawking radiation rests on the validity of a more stringent {\em global} criterion for thermalization. How long does it take for the quantum information deposited by Alice to become thoroughly mixed with the black hole's microscopic degrees of freedom?

Since this is really a question about strongly-coupled quantum gravity, we don't have the tools to answer it definitively, but we can try to make a plausible guess. For this purpose, we envision the black hole's qubits as uniformly distributed over the ``stretched horizon'' \cite{complement,susskind-book}. The stretched horizon is located about one Planck unit of proper distance above the actual global horizon; here static observers have a proper acceleration of order one in Planck units, and therefore are surrounded by a bath of Unruh radiation with temperature of order one. (Quanta from this bath that are directed nearly radially outward can escape to infinity as Hawking radiation, with a redshifted temperature of order $1/r_S$.) 

According to the hypothesis of ``black hole complementarity'' \cite{complement,susskind-book}, observers who remain outside the black hole may legitimately regard the stretched horizon as a repository that stores quantum information absorbed by the black hole, even though the strongly-coupled ``Planckian soup'' is invisible to freely falling observers who pass through the horizon. Because of the high temperature of the Planckian soup at the stretched horizon, we need a thorough grasp of quantum gravity to understand its detailed dynamics. 

The stretched horizon has some dissipative properties that can be analyzed classically \cite{membrane}. For example, if electric charge is deposited in the black hole, currents flow on the stretched horizon to redistribute the charge, which are damped by the stretched horizon's internal resistance. The local charge density decays like $\exp(-t/r_S)$, where $t$ is the Schwarzschild time, and correspondingly the proper area of a droplet of charge on the stretched horizon grows like $\exp(t/r_S)$. The droplet envelops the stretched horizon, and the charge distribution reaches a new steady state, after a Schwarzschild time of order $r_S\log r_S$. From the perspective of static observers hovering above the stretched horizon, the spreading of charge is very fast, reaching a proper distance of order $r_S$ in a time of order $\log r_S$ as measured by their clocks; yet these observers are unable to manipulate the flow of charge to achieve superluminal communication.

In its classical realization, the superluminal spreading of electric charge at the stretched horizon is really a kinematic effect --- it arises because the static observers in the Schwarzschild geometry separate with constant proper acceleration from freely falling charged particles (beneath the stretched horizon) that are plunging toward the black hole's global event horizon. One feels hesitant about drawing any far-reaching dynamic conclusions based on this kinematic spreading. But on the other hand a complete physical description of physics at the stretched horizon should accommodate some kind of rapidly mixing dynamics, making it hard to distinguish between effects that arise from strong local interactions and effects that are merely kinematic. Thus one is tempted to postulate that the exponential spreading is exhibited not just by the black hole's classical hair, but also by its ``quantum hair;''  i.e., the quantum information that it encodes. 

To the observers anchored at the stretched horizon, not just the local charge density but also other local disturbances of the thermal bath decay like $\exp(-t_{\rm stretch})$, where $t_{\rm stretch}$ is the time in Planck units as measured by static clocks at the stretched horizon. If we (somewhat fancifully) envision ``information'' deposited in the black hole as a locally conserved fluid, the $\exp(-t_{\rm stretch})$ decay of the local information density suggests that the droplet of information, just like the droplet of charge, expands exponentially and becomes nearly uniformly distributed over the stretched horizon in a time $t_{\rm stretch}=O(\log r_S)$, corresponding to Schwarzschild time $t=O(r_S\log r_S)$. This picture leads us to suggest that the black hole's global (Schwarzschild) thermalization time is $O(r_S\log r_S)$. 

In a sense this suggestion (also put forward in \cite{susskind-book}) only deepens the information puzzle, as it poses the challenge of reconciling rapid spreading of quantum information on the stretched horizon with the black hole's causal structure. On the other hand, rapid distribution of many-particle quantum entanglement need not imply superluminal signaling. Furthermore, the black hole complementarity viewpoint already postulates that the semiclassical causal structure of a black hole must be highly misleading in some respects; the superluminal spreading is a relatively gentle extension of a viewpoint we are already taking for granted.

If an observer who is a distance of order $r_S$ from the horizon drops a freely falling quantum memory into the black hole, it takes Schwarzschild time of order $r_S\log r_S$ for the memory to reach the stretched horizon, and conversely it takes Schwarzschild time of order $r_S\log r_S$ for quanta emitted from the stretch horizon to reach the observer's detector. Therefore, a thermalization time of order $r_S\log r_S$ is compatible with our central metaphor --- the black hole is a ``mirror'' in the sense that it returns Alice's information in a time comparable to the time it takes the information to fall into the black hole.

If one, more conservatively, insists that the spreading ``droplet of information'' remains confined to the forward light cone, then the time required for global thermalization is $t_{\rm stretch}=\Omega(r_S)$, corresponding to Schwarzschild time $t=\Omega(r_S^2)$. (Here the ``big-Omega'' notation indicates a {\em lower bound} on the thermalization time for asymptotically large $r_S$, up to a multiplicative constant.) In that event, since once global thermalization is achieved $k$ qubits can escape in Schwarzschild time $t=O(kr_S)$, one may expect that the Schwarzschild time for a constant number of qubits to escape from the black hole is $t=O(r_S^p\log^q r_S)$, where $1 < p < 2$. Here the exponents $p$ and $q$ are sensitive to the details of the model of thermalization. 

Of our two main conclusions in Sec.~\ref{sec:quantum}, the conclusion that Alice's qubits bounce right back (if the black hole has already radiated away more than half of its initial entropy) really does require rapid thermalization, i.e., thermalization in Schwarzschild time $t=O(r_S\log r_S)$. But the other conclusion, that previously absorbed quantum information is released quickly when the evaporation reaches the half-way point, still applies even if thermalization takes much longer. For the second conclusion, we only need thermalization to be fast compared to evaporation, which occurs in a Schwarzschild time of order $r_S^3$ rather than $r_S\log r_S$.

\section{Efficiency}
\label{sec:efficiency}

In Sec.~\ref{sec:quantum}, our analysis of the quantum model of a black hole was, as a computer scientist might put it derisively, ``merely information theoretic.'' We in effect assumed without justification that the unitary transformation arising from the internal dynamics of an $n$-qubit black hole is typical with respect to the Haar measure on $U(2^n)$ (and we also ignored the complexity of Bob's decoding algorithm). The vague picture offered in Sec.~\ref{sec:thermalization}, depicting thermalization as the rapid spreading of a droplet, does not by itself provide much support for describing the black hole's dynamics as a random unitary transformation. Can we defend this assumption?

In accord with the black hole complementarity hypothesis, let us suppose that the internal dynamics of the black hole is governed by a strongly mixing local Hamiltonian acting on $n$ qubits that are uniformly distributed over the black hole's stretched horizon \cite{complement,susskind-book}. Here $n$ (up to the factor $\frac{1}{4}\log_2 e$) is the area of the horizon in Planck units, and the Schwarzschild radius is $r_S=O(\sqrt{n})$ in Planck units. Though we don't understand quantum gravity well enough to analyze the internal dynamics in detail,  we might nevertheless hope to make a few cogent general observations.

We find it helpful to envision this dynamics as a local quantum circuit, where in each unit of time two-qubit unitary transformations (``quantum gates'') are applied in parallel to about $n/2$ pairs of neighboring qubits. It seems natural to choose the unit of time to be the Planck time, but according to whose clock? This is a subtle question, because, due to the gravitational redshift, the clocks of static observers hovering above the stretched horizon run much slower (by a factor of order $r_S$ in Planck units) than the clocks of observers residing far from the horizon. As already noted in Sec.~\ref{sec:thermalization}, there are physical processes on the stretched horizon (like the distribution of electric charge) that can act over a proper distance of order $r_S$ in a Schwarzschild time of order $r_S$. If we wish to accommodate such processes in our circuit model, we may take the unit of time to be a Planck unit of Schwarzschild time. This model of the local dynamics on the stretched horizon is very naive, and in particular the model does not attempt to address the challenge of reconciling rapid spreading of information with causality. In any case, no matter how we prefer to map circuit time to physical time, it is useful to characterize the circuit complexity of global thermalization. 

In our model, after a Schwarzschild time $t$, the unitary transformation acting on the black hole's internal state is realized by a circuit that has a number of time steps (the circuit's ``depth'') of order $t$ (where $t$ is expressed in Planck units) and a total number of (local) quantum gates (the circuit's ``size'') of order $nt$. Unfortunately, if $t$ is a polynomial in $n$, then Haar-random unitary transformations cannot be generated with circuits of this size. In fact, because the volume of $U(2^n)$ is exponentially large in $2^n$, reaching a typical unitary transformation requires a circuit whose size is exponential in $n$. 

However, much smaller circuits can be good encoders for quantum error-correcting codes. In particular, the capacity of the quantum erasure channel is achieved by typical {\em stabilizer codes}, which have encoding circuits of size $O(n^2)$ \cite{gottesman-thesis}. Furthermore, good approximate encoding circuits can be even smaller. If we replace the integral over all unitary transformations by an average over the elements of the $\epsilon$-approximate unitary 2-design ${\cal K}$ constructed in \cite{cleve}, then eq.~(\ref{quantum-out}) is modified to become
\begin{eqnarray}
\label{design-out}
\frac{1}{|{\cal K}|}\sum_{k\in{\cal K}} \left\| \sigma^{NB'}(V_k^B)-\sigma^{N}(V_k^B)\otimes \sigma^{B'}_{\rm max}\right\|_1^2 ~\le ~ 2^{-2(s-k)} +\epsilon + O(2^{-n})~.
\end{eqnarray}
Each element of ${\cal K}$ can be realized by a quantum circuit that has size $O(n\log(1/\epsilon))$ and depth $O(\log n\log (1/\epsilon))$ \cite{cleve,dankert}. It seems reasonable to expect that random quantum circuits built from two-qubit gates would encode at least as efficiently as the approximate 2-designs that have been explicitly constructed, and preliminary results from numerical experiments are consistent with this expectation \cite{anderson-sim}. 

This estimate of the circuit size and depth does not take into account any geometric locality constraints.  If the $n$ qubits are uniformly distributed on a two-dimensional sphere with radius of order $\sqrt{n}$, then  the additional cost of using local gates should be at worst a blowup in size and depth by a factor of order $\sqrt{n}$, since the information stored in a qubit can propagate a distance $\sqrt{n}$ when $\sqrt{n}$ local gates are applied in succession. Therefore, there exist good approximate local encoding circuits for the quantum erasure channel that have size $O(n^{3/2}\log(1/\epsilon))$ and depth $O(n^{1/2}\log n\log(1/\epsilon))$. Again, we expect that random local circuits of this size and depth would do at least as well.

Therefore, if we model the internal dynamics of the stretched horizon by a random local quantum circuit, we conclude that the global thermalization of the black hole is achieved by a circuit whose depth is $O(r_S\log r_S \log(1/\epsilon))$. By ``global thermalization'' we mean such that eq.~(\ref{design-out}) holds, so that Bob can approximately decode Alice's quantum memory with a loss of fidelity of order $\epsilon$. Modeling the internal dynamics of a physical system by a small random quantum circuit is far more defensible than modeling it by a unitary chosen uniformly with respect to Haar measure, because a small circuit can  be realized by physically plausible local dynamics. It therefore seems reasonable to suggest that the dynamics of a black hole could efficiently generate good quantum error-correcting codes for the quantum erasure channel. Indeed, if each computational step occurs in one Planck unit of Schwarzschild time, then the thermalization time $t=O(r_S\log r_S)$ scales with $r_S$ as in our crude estimate in Sec.~\ref{sec:thermalization}.

Decoding, however, can be a much harder computational problem than encoding. Black hole evaporation maps Alice's $k$ qubits to a highly nonlocal subsystem $M'$ of $RE$. Bob's decoding operation maps $M'$ back to a compact system $\hat M$ localized in his laboratory, returning the qubits to a usable form. Even if we are willing to grant Bob the power to collect all of the Hawking radiation emitted by the black hole, Bob might not be able to recover Alice's $k$-qubit message by means of an efficient quantum computation (the resources needed to perform Bob's computation might grow faster than any power of $k$). 

Actually, if a stabilizer code is used to protect entanglement-assisted quantum communication through the erasure channel, then Bob's decoding computation is easy. Bob can replace each erased qubit by the standard state $|0\rangle$, and then measure the code's check operators. With high probability, there is a unique Pauli operator acting on the erased qubits that restores Bob's state to the code space, and the recovery operation can be efficiently computed using linear algebra. However, this recovery procedure exploits the special structure of stabilizer codes, and would not work for typical non-stabilizer codes. If thermalization is rapid, then the quantum codes realized by evaporating black holes have small encoding circuits and therefore they too are rather special, but they are not stabilizer codes and we do not know whether they are efficiently decodable. Conceivably, the decoding is hard not only for $n$ large with $k/n$ fixed, but also for $n$ large with $k$ fixed. If Bob's decoding problem is intractable, then the physicists of our highly advanced civilization may be hard pressed to test the hypothesis that black holes reradiate quantum information quickly. 

For that matter, one might question the testability of the claim that black holes really preserve quantum information. We emphasize, though, that even if Bob's decoding problem is hopelessly intractable, the physicists of our highly advanced civilization, using only ``polynomial resources,'' should be able to test the hypothesis that black holes process quantum information without destroying it. Once these physicists understand the laws of black hole dynamics well enough, they should be able, using their quantum computers, to simulate the formation and complete evaporation of a black hole. Through a ``swap test'' \cite{swap}, they could compare the output of their quantum simulation with the observed product of the black hole evaporation, and so verify that the dynamical black hole is behaving as their theory predicts. The ``feasibility'' (in principle) of such a test boosts our confidence that the assertion that black holes preserve information has a well formulated operational meaning. Unfortunately, the swap test works for comparing pure states, not mixed states; it can't be used to compare the simulation with experiment in the case of a partially evaporated black hole. 

\section{Are black holes quantum cloners?}
\label{sec:cloning}

Our analysis has been premised on the assumption that black hole evaporation respects unitarity, and we have adopted the black hole complementarity viewpoint to describe the process. We have been led to the conclusion that a black hole (whose evaporation is past the half-way point) is really a sort of ``information mirror'' that quickly returns the quantum information it receives. Now we should reassess whether our assumptions are consistent. 

We note that the geometry of the evaporating black hole contains spacelike surfaces that are crossed  both by Alice's infalling quantum memory (inside the event horizon), and by the outgoing Hawking radiation that Bob decodes (outside the horizon). Therefore, if Bob can decode successfully, then Alice's quantum information has been ``cloned'' in the outgoing radiation. But cloning of arbitrary quantum states is inconsistent with the linearity of quantum mechanics \cite{no-cloning,dieks}. We seem to be stuck with a difficult choice: either Alice's information is not reemitted, or the black hole is a quantum cloning machine! Either way, the foundations of quantum theory need revision. This appears to be a powerful argument in favor of information loss \cite{preskill-destroy}. 

The hypothesis of black hole complementarity was proposed as a way to avoid this quandary \cite{complement,susskind-book}: one chooses not to be bothered by quantum cloning if it occurs where no one can ever find out. According to this philosophy, we may accept for now that Alice (if she falls into the black hole) and Bob (if he stays outside) have sharply contrasting descriptions of the same physical process. Eventually we hope to be able to reconcile Alice's and Bob's contrasting viewpoints, but finding that more complete global description must be postponed until we acquire a deeper understanding of quantum gravity. In the meantime, we are entitled to insist that Alice's and Bob's descriptions are both compatible with the standard principles of quantum mechanics.

Therefore we should ask: if Alice's quantum state persists behind the horizon and that state is also encoded in the outgoing Hawking radiation received by Bob, can Alice or Bob {\em verify} the cloning? Once Alice falls through the horizon, she won't be able to compare notes with Bob if he stays outside the black hole. To have any hope of verifying the cloning, Bob needs to remain outside until he has retrieved Alice's qubits from the Hawking radiation, and then dive into the black hole seeking confirmation that both he and Alice have high-fidelity copies of Alice's original quantum state. However, as long as the black hole retains Alice's qubits long enough before reemitting them, then we are unable to say (based on our current understanding of quantum gravity) whether Bob will succeed or not \cite{susskind-gedanken,preskill-unpublished}. As best we can tell, then, the black hole complementarity hypothesis seems to be self-consistent, provided the black hole retains quantum information for a sufficiently long time.

\begin{figure}
\begin{center}
\leavevmode
\epsfxsize=4.5in
\epsfbox{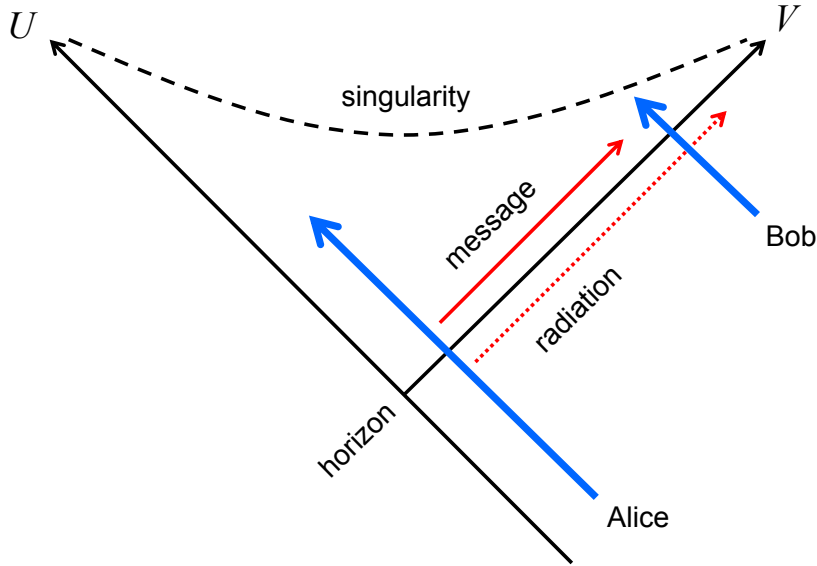}
\end{center}
\caption{Alice and Bob test whether an evaporating black hole clones quantum information. Alice, carrying her quantum memory, drops into the black hole. Bob recovers the content of Alice's memory from the Hawking radiation, and then enters the black hole, too. Alice sends her qubits to Bob, and Bob verifies that cloning has occurred.}
\label{fig:cloning}
\end{figure}

To analyze this thought experiment, it is convenient to describe the Schwarzschild black hole using the Kruskal null coordinates $U,V$ as depicted in Fig.~\ref{fig:cloning}. These coordinates are related to the Schwarzschild coordinates $r,t$ by
\begin{eqnarray}
U=-e^{(r_*-t)/2r_S}~,\quad V=e^{(r_*+t)/2r_S}~,\quad {\rm where} \quad r_*=r+r_S\ln[(r-r_S)/r_S]~.
\end{eqnarray}
In terms of these coordinates, the black hole's event horizon is at $U=0$, and the curvature singularity occurs at $UV=1$. Therefore, if Bob crosses the horizon at $V=V_{\rm Bob}$, then he reaches the singularity at $U\le V_{\rm Bob}^{-1}$. But if Alice falls freely, the proper time she experiences between crossing the horizon at $V=V_{\rm Alice}$ and reaching $U=V_{\rm Bob}^{-1}$ is 
\begin{eqnarray}
\tau_{\rm Alice}= Cr_S\left({V_{\rm Alice}}/{V_{\rm Bob}}\right)~,
\end{eqnarray}
where $C$ is a numerical constant that depends on Alice's initial data ($C=1/e\approx .368$ if Alice falls from rest starting at $r=\infty$). In terms of Schwarzschild time, Bob's fall into the black hole is delayed relative to Alice's by $\Delta t$, where $V_{\rm Bob}/V_{\rm Alice}= e^{\Delta t/2r_S}$, and therefore
\begin{eqnarray}
\tau_{\rm Alice}= C  r_S\exp(-\Delta t/2r_S)~.
\end{eqnarray}
Thus Alice's proper time $\tau_{\rm Alice}$ is of order the Planck time or shorter if Bob's entry into the black hole is delayed by $\Delta t = r_S\log r_S$ or longer. The conclusion does not change substantially even if Alice rides a rocket, as long as her proper acceleration is small in Planck units. 

If Alice, after crossing the horizon, has less than a Planck time to communicate with Bob about the status of her qubits, then she is required to send her message to Bob using super-Planckian frequencies. And with our incomplete understanding of strongly-coupled quantum gravity, we lack the tools to analyze the emission, transmission, and reception of such signals. On the other hand, if Alice's proper time were long compared to the Planck time, then in principle she could send her qubits to Bob (or send a record of her measurement outcomes), allowing Bob to verify the cloning, and the verification could be analyzed using controlled semiclassical approximations. We conclude that, if verifiable cloning does not occur, then the black hole must retain Alice's information for a Schwarzschild time interval $\Delta t=\Omega(r_S\log r_S$) \cite{susskind-gedanken,preskill-unpublished}. (The ``big-Omega'' notation indicates that $r_S\log r_S$ is a {\em lower bound} on the information retention time for asymptotically large $r_S$, up to a multiplicative constant.) This finding is just barely compatible with the information retention time estimated in Sec.~\ref{sec:thermalization} and Sec.~\ref{sec:efficiency}.

We noted in Sec.~\ref{sec:thermalization} that it already takes Schwarzschild time $O(r_S\log r_S)$ for the Hawking radiation to climb from the stretched horizon to Bob's detectors, if Bob waits at a position that is a proper distance $O(r_S)$ from the horizon. In principle, though, Bob's  ``decoding sphere'' could be located at a height where the black hole's thermal atmosphere has a constant temperature (independent of $r_S$), At this height, Bob's clock runs slower than Schwarzschild time by a factor $O(r_S)$, and the outgoing radiation propagates from the stretched horizon to Bob in $O(1)$ time. From the requirement that cloning is unverifiable, we infer that by Bob's clock it should take time $\Omega(\log r_S)$ for Alice's quantum information to reappear in the thermal atmosphere. This version of the ``no-cloning'' argument usefully differentiates between the time delay due to the climb from the stretched horizon and the time delay due to the thermalization of the stretched horizon.

A serious weakness in the ``no-cloning'' argument is that we have assumed that, once Alice's information is received by Bob, he can decode the information instantaneously. In fact, as emphasized in Sec.~\ref{sec:efficiency}, Bob's decoding map is a complex quantum operation acting on $O(r_S^2)$ qubits. One could plausibly argue that fundamental laws of physics prevent Bob from performing the decoding operation fast enough for cloning to be verified behind the horizon, no matter how quickly Alice's qubits become encoded in the Hawking radiation. Nevertheless, we find it intriguing that our estimate of the information retention time in Sec.~\ref{sec:efficiency} is of the same order as the ``no-cloning'' limit that can be inferred if we neglect Bob's decoding time.

\section{Conclusions}
The purpose of this article is largely pedagogical. On the one hand, many physicists who are interested in the quantum behavior of black holes are familiar with Don Page's observations concerning the entanglement entropy of an evaporating black hole \cite{page-entropy}. On the other hand, most quantum information theorists know about the quantum capacity of the quantum erasure channel \cite{thapliyal,quantum-erasure}. Here we point out that by invoking the latter we can refine and extend the former. We have tried to convey our main points without relying too much on information-theoretic jargon, in the hope of reaching a broad audience among those who are intrigued by these issues.

The essential tool that allows us to go further than Page is the theory of quantum error correction, which was developed after \cite{page-entropy} was written. In fact there is a delicious irony at the core of our analysis. In Sec.~\ref{sec:quantum}, we modeled the internal dynamics of the black hole with a randomly selected unitary transformation, presuming that black holes, though they do not destroy information, are nearly optimal thermalizers --- they rapidly convert information to a form that is very difficult to access in practice. However, a random unitary is also a nearly optimal encoder that achieves the quantum capacity for the quantum erasure channel. The dynamics that conceals quantum information effectively in practice, by encoding it in a highly nonlocal manner, also at the same time protects the information from damage as a matter of principle. 

It is not fully realistic to model the internal dynamics with a random unitary transformation, because typical unitary transformations can be generated only by quantum circuits of exponential size. However, in Sec.~\ref{sec:efficiency} we invoked recent results concerning efficient approximate encoding circuits for quantum error-correcting codes \cite{cleve,dankert}, indicating that plausible local dynamics really can thermalize quantum information quickly. We argued in Sec.~\ref{sec:cloning} that our estimate of a black hole's information retention time is just barely compatible with the principle that black hole evaporation must not realize verifiable quantum cloning.

Finally, what, if anything, are the broader implications of our observations for the black hole information puzzle? We're not sure, but we find it intriguing that the encoder for a good quantum error-correcting code need not necessarily be unitary. Perhaps, then, there could be an interesting middle ground between unitary black hole dynamics and full-blown information loss --- even if some of the information deposited in a black hole remains inaccessible forever, quantum information encoded in small subsystems might escape in the Hawking radiation with relatively little damage.

One possible model is to suppose that at any given time the black hole system $B'$ actually consists of two parts, the ``accessible'' system $B_{\rm acc}$, for which we might suppose that $\log |B_{\rm acc}|$ is the Bekenstein-Hawking entropy, and an inaccessible system $B_{\rm gone}$ which has been lost forever. For the information to be revealed to Bob, we require that correlations are destroyed not just between the reference system $N$ and $B_{\rm acc}$, but rather between $N$ and $B'=B_{\rm acc}B_{\rm gone}$. As the radiation leaks out, $B_{\rm acc}$ shrinks, but $B_{\rm gone}$ may grow, and there is a competition between these two effects. If $R$ is radiated away, then the effective dimension of the discarded system is not $|R|$ but rather $|R|/|B_{\rm gone}|$; Alice's quantum information is revealed when this effective dimension becomes larger than $|N|$. Quantum information encoded in a sufficiently small subsystem eventually escapes if the qubits are emitted in the radiation faster than they are destroyed in the black hole's interior, but the rate of escape is suppressed to a degree that depends on the rate of destruction. 

Since information loss, once allowed, tends to be highly infectious, it is difficult to formulate deformations of quantum mechanics that incorporate a small amount of information loss, yet are compatible with low-energy experimental constraints \cite{banks}. In the scheme we have just outlined, when a pure quantum state undergoes gravitational collapse to form a black hole and then evaporates completely, some of the information encoded in the initial quantum state really is destroyed, but the information encoded in any sufficiently small subsystem survives nearly unscathed. It might be fruitful to investigate how easily this type of {\em partial} information loss can be reconciled with the experimental successes of quantum mechanics.

\vskip .5cm
\begin{center} {\bf Acknowledgments} \end{center}

We are grateful for the hospitality of the Perimeter Institute, where we had the good fortune to share an office, and JP thanks PH for letting him use the comfortable chair. We also thank Ashton Anderson, Hilary Carteret, Daniel Gottesman,  Dennis Kretschmann, Seth Lloyd, Prakash Panangaden, David Poulin, Renato Renner, Lenny Susskind, Kip Thorne, Bill Unruh, Andreas Winter, Jon Yard, and the participants in the 2007 McGill-Bellairs Quantum Information Workshop for helpful suggestions. This research is supported in part by the Canada Research Chairs program, the Sloan Foundation, CIFAR, FQRNT, MITACS, NSERC, DoE under Grant No. DE-FG03-92-ER40701, NSF under Grant No. PHY-0456720, and NSA under ARO Contract No. W911NF-05-1-0294.


\begin{thebibliography}{99}
\bibitem{hawk2} S.~W. Hawking, ``Breakdown of predictability in gravitational collapse,'' Phys. Rev. D {\bf 14}, 2460 (1976).
\bibitem{strominger-vafa} A Strominger and C. Vafa, ``Microscopic origin of the Bekenstein-Hawking entropy,'' Phys. Lett. B {\bf 379}, 99-104 (1996), arXiv:hep-th/9601029.
\bibitem{ads-cft} J. Maldacena, ``The large-$N$ limit of superconformal field theories and supergravity,'' Adv. Theor. Math. Phys. {\bf 2}, 231 (1998), arXiv:hep-th/9711200.
\bibitem{bennett-shor} C.~H. Bennett and P.~W. Shor, ``Quantum information theory,'' IEEE Trans. Information Theory {\bf 44}, 2724-2742 (1998).
\bibitem{thapliyal} C.~H. Bennett, P.~W. Shor, J.~A. Smolin, and A.~V. Thapliyal, ``Entanglement-assisted classical capacity of noisy quantum channels,'' Phys. Rev. Lett. {\bf 83}, 3081 (1999), arXiv:quant-ph/9904023.
\bibitem{cleve} C. Dankert, R. Cleve, J. Emerson, and E. Livine, ``Exact and approximate unitary 2-designs: constructions and applications,'' arXiv:quant-ph/0606161 (2006).
\bibitem{dankert} C. Dankert, ``Efficient simulation of random quantum states and operators,''  arXiv:quant-ph/0512217 (2005).
\bibitem{complement} L. Susskind, L. Thorlacius, and J. Uglum, ``The stretched horizon and black hole complementarity,'' Phys. Rev. {\bf 48}, 3743 (1993), arXiv:hep-th/9306069.
\bibitem{susskind-book} L. Susskind and J.~S. Lindesay, {\em Black Holes, Information, and the String Theory Revolution: The Holographic Universe} (Singapore, World Scientific, 2005).
\bibitem{cover} T.~M. Cover and J.~A. Thomas, {\em Elements of Information Theory} (New York, Wiley, 1991).
\bibitem{barnum} H. Barnum, E. Knill, and M.~A. Nielsen, ``On quantum fidelities and channel capacities,'' IEEE Trans. Info. Theor. {\bf 46}, 1317-1329 (2000), arXiv:quant-ph/9809010.
\bibitem{page-entropy} D.~N Page, ``Average entropy of a subsystem,'' Phys. Rev. Lett. {\bf 71}, 1291 (1993), arXiv:gr-qc/9305007; D.~N. Page, ``Black hole information,'' in {\em Proceedings of the 5th Canadian Conference on General Relativity and Relativistic Astrophysics}, ed. R.~B. Mann and R.~G. McLenaghan (1994), arXiv:hep-th/9305040.
\bibitem{lubkin} E. Lubkin, ``Entropy of an $n$-system from its correlation with a $k$-reservoir, J. Math. Phys. {\bf 19}, 1028 (1978).
\bibitem{lloyd} S. Lloyd, ``Black holes, demons, and loss of coherence: how complex systems get information and what they do with it,'' Rockefeller University Ph.D. thesis (1988).
\bibitem{abeyesinghe} A. Abeyesinghe, I. Devetak, P. Hayden, and A. Winter, ``The mother of all protocols: Restructuring quantum information's family tree,'' arXiv:quant-ph/0606225 (2006).
\bibitem{yard} P. Hayden, M. Horodecki, A. Winter and J. Yard, ``A decoupling approach to the quantum capacity,'' arXiv:quant-ph/0702005 (2007).
\bibitem{fuchs} C.~A. Fuchs, ``Distinguishability and accessible information in quantum theory,'' arXiv:quant-ph/9601020 (1996). 
\bibitem{quantum-erasure} C.~H. Bennett, D.~P. DiVincenzo, and J.~A. Smolin, ``Capacities of quantum erasure channels,'' Phys. Rev. Lett. {\bf 78}, 3217 (1997), arXiv:quant-ph/9701015. 
\bibitem{mehta} M.~L. Mehta, {\em Random Matrices} (Amsterdam, Elsevier, 2004). 
\bibitem{anderson} A. Anderson and P. Hayden, ``Global thermalization and random matrix
theory,'' in preparation (2007).
\bibitem{horowitz} G.~T. Horowitz and V.~E. Hubeny,``Quasinormal modes of AdS black holes and the approach to thermal equilibrium,'' Phys.Rev. D {\bf 62}, 024027 (2000), arXiv:hep-th/9909056.
\bibitem{membrane} K.~S. Thorne, R.~H. Price, and D.~A. Macdonald (eds.), {\em Black Holes: The Membrane Paradigm} (New Haven, Yale, 1986).
\bibitem{gottesman-thesis} D. Gottesman, ``Stabilizer codes and quantum error correction,'' arXiv:quant-ph/9705052 (1997). 
\bibitem{anderson-sim} A. Anderson and P. Hayden, unpublished (2007).
\bibitem{swap}  H. Buhrman, R. Cleve, J. Watrous, and R. de Wolf, ``Quantum fingerprinting,'' arXiv:quant-ph/0102001 (2001).
\bibitem{no-cloning} W.~K. Wootters and W.~H. Zurek, ``A single quantum cannot be cloned,'' Nature {\bf 299}, 802-803 (1982).
\bibitem{dieks} D. Dieks, ``Communication by EPR devices,'' Physics Letters A {\bf 92}, 271-272 (1982).
\bibitem{preskill-destroy} J. Preskill, ``Do black holes destroy information?'' hep-th/9209058 (1992).
\bibitem{susskind-gedanken} L. Susskind and L. Thorlacius, ``Gedanken experiments involving black holes,'' Phys. Rev. D {\bf 49}, 966-974 (1994), hep-th/9308100.
\bibitem{preskill-unpublished} J. Preskill, unpublished (1993).
\bibitem{banks} T. Banks, M.~E. Peskin, and L. Susskind, ``Difficulties for the evolution of pure states into mixed states,'' Nucl. Phys. B {\bf 244}, 125 (1984).

\end{thebibliography}
\end{document}